\documentclass[12pt, letterpaper]{JHEP3}
\usepackage{amssymb, amsmath, amsopn, amsthm}
\usepackage{epsfig}
\usepackage{psfrag}
\usepackage{subfigure}

\newcommand{\eref}[1]{(\ref{#1})}

\newcommand{\fref}[1]{Figure~\ref{#1}}
\newcommand{\cref}[1]{Chapter~\ref{#1}}
\newcommand{\beq}{\begin{equation}}
\newcommand{\eeq}{\end{equation}}
\newcommand{\ba}{\begin{array}}
\newcommand{\ea}{\end{array}}
\newcommand{\bcenter}{\begin{center}}
\newcommand{\ecenter}{\end{center}}

\def\IB{\relax\hbox{$\inbar\kern-.3em{\rm B}$}}
\def\IC{\relax\hbox{$\inbar\kern-.3em{\rm C}$}}
\def\ID{\relax\hbox{$\inbar\kern-.3em{\rm D}$}}
\def\IE{\relax\hbox{$\inbar\kern-.3em{\rm E}$}}
\def\IF{\relax\hbox{$\inbar\kern-.3em{\rm F}$}}
\def\IG{\relax\hbox{$\inbar\kern-.3em{\rm G}$}}
\def\IGa{\relax\hbox{${\rm I}\kern-.18em\Gamma$}}
\def\IH{\relax{\rm I\kern-.18em H}}
\def\IK{\relax{\rm I\kern-.18em K}}
\def\IL{\relax{\rm I\kern-.18em L}}
\def\IP{\relax{\rm I\kern-.18em P}}
\def\IR{\relax{\rm I\kern-.18em R}}
\def\IZ{\relax\ifmmode\mathchoice
{\hbox{\cmss Z\kern-.4em Z}}{\hbox{\cmss Z\kern-.4em Z}}
{\lower.9pt\hbox{\cmsss Z\kern-.4em Z}}
{\lower1.2pt\hbox{\cmsss Z\kern-.4em Z}}\else{\cmss Z\kern-.4em Z}\fi}
\def\II{\relax{\rm I\kern-.18em I}}

\def\sCC{{\kern 0.27em\vrule height1.45ex width0.03em depth0em
          \kern-0.30em\rm C}}
\def\C{{\mathchoice
  {\sCC}
  {\sCC}
  {\kern 0.225em \vrule height1.05ex width0.025em depth0em \kern-0.25em \rm C}
  {\kern 0.180em \vrule height0.78ex width0.02em depth0em \kern-0.2em \rm C}
        }}
\def\sHH{{\rm I\kern-.16em{}H}}
\def\H{{\mathchoice
  {\sHH}
  {\sHH}
  {\rm I\kern-.13em{}H}
  {\rm I\kern-.13em{}H} }}
\def\sNN{{\rm I\kern-.16em{}N}}
\def\N{{\mathchoice
  {\sNN}
  {\sNN}
  {\rm I\kern-.12em{}N}
  {\rm I\kern-.10em{}N} }}
\def\sPP{{\rm I\kern-.16em{}P}}
\def\P{{\mathchoice
  {\sPP}
  {\sPP}
  {\rm I\kern-.12em{}P}
  {\rm I\kern-.10em{}P} }}
\def\sQQ{{\kern 0.27em \vrule height1.45ex width0.03em depth0em
          \kern-0.30em \rm Q}}
\def\Q{{\mathchoice
        {\sQQ}
        {\sQQ}
  {\kern 0.225em \vrule height1.05ex width0.025em depth0em \kern-0.25em \rm Q}
  {\kern 0.180em \vrule height0.78ex width0.020em depth0em \kern-0.20em \rm Q}
        }}
\def\sRR{{\rm I\kern-0.16em{}R}}
\def\R{{\mathchoice
  {\sRR}
  {\sRR}
  {\rm I\kern-0.12em{}R}
  {\rm I\kern-0.10em{}R} }}
\def\sZZ{{\rm Z\kern-0.32em{}Z}}
\def\Z{{\mathchoice
  {\sZZ}
  {\sZZ} 
  {\rm Z\kern-0.3em{}Z}     
  {\rm Z\kern-0.25em{}Z} }}  
\def\ZZZ{{\rm Z\kern-0.24em{}Z}}
\def\sII{{\rm I\kern-0.16em{}I}}
\def\I{{\mathchoice
  {\sII}
  {\sII}
  {\rm I\kern-0.12em{}I}
  {\rm I\kern-0.10em{}I} }}

\def\inbar{\,\vrule height1.5ex width.4pt depth0pt}
\font\cmss=cmss10 \font\cmsss=cmss10 at 7pt

\def\smiley{\hbox{\large$\bigcirc$\hspace{-0.80em}\raise.2ex
\hbox{$\cdot\cdot$}\kern-.61em\lower.2ex\hbox{\scriptsize$\smile$}}\ }
\def\frowny{\hbox{\large$\bigcirc$\hspace{-0.80em}\raise.2ex
\hbox{$\cdot\cdot$}\kern-.635em\lower.2ex\hbox{\scriptsize$\frown$}}\ }

\def\I{{\rlap{1} \hskip 1.6pt \hbox{1}}}

\makeatletter
\let\hangafter\@hangfrom
\makeatother

\title{Metastable vacua and D-branes at the conifold}

\author{Riccardo Argurio$^1$, Matteo Bertolini$^2$,
Sebasti\'an Franco$^3$ and Shamit Kachru$^4$

\\
~\\

${}^1$Physique Th\'eorique et Math\'ematique and International Solvay Institutes \\
Universit\'e Libre de Bruxelles, C.P. 231, 1050 Bruxelles, Belgium \\ \vspace{0.3cm}
${}^2$SISSA/ISAS and INFN - Sezione di Trieste \\
Via Beirut 2; I 34014 Trieste, Italy \\ \vspace{0.3cm}
${}^3$Joseph Henry Laboratories, Princeton University \\
Princeton, NJ 08544, USA \\ \vspace{0.3cm}
${}^4$Department of Physics and SLAC, Stanford University \\
Stanford, CA 94305 USA \\

\email{rargurio@ulb.ac.be, bertmat@sissa.it, sfranco@feynman.princeton.edu,
skachru@stanford.edu}\\
}

\abstract{
We consider quiver gauge theories arising on D-branes at simple Calabi-Yau
singularities (quotients of the conifold).  These theories  
have metastable
supersymmetry breaking vacua.
The field theoretic mechanism is basically
the one exhibited by the examples of Intriligator, Seiberg and Shih in SUSY QCD.
In a dual description, 
the SUSY breaking is captured by the presence of anti-branes.
In comparison to our earlier related work, the main improvements of the
present construction are that we can reach the free magnetic range of
the SUSY QCD theory where the existence of the metastable vacua is on
firm footing, and we can see explicitly how the small masses for the quark
flavors (necessary to the existence of the SUSY breaking vacua) are
dynamically stabilized. One crucial mass term is generated
by a stringy instanton. Finally, our models
naturally incorporate R-symmetry breaking in the
non-supersymmetric vacuum, in a way similar to the examples of
Kitano, Ooguri and Ookouchi.}

\preprint{SISSA-15/2007/EP \\ PUPT-2228 \\SLAC-PUB-12383}

\def\be{\begin{equation}}
\def\ee{\end{equation}}
\def\bea{\begin{eqnarray}}
\def\eea{\end{eqnarray}}

\newcommand{\cO}{\mathcal{O}}

\newcommand{\bC}{\mathbb{C}}

\newcommand{\bZ}{\mathbb{Z}}

\begin{document}
\tableofcontents

\section{Introduction}

There has been a renaissance in the study of metastable
supersymmetry breaking vacua
in string theory and field theory in the past several years.
Several of the examples are stringy constructions which
involve brane dynamics
and/or fluxes in
a nontrivial gravitational background \cite{KachruMcGreevy, VafaLargeN,
KachruPearsonVerlinde,
ABSV, VerlindeMonopole, HeckmanVafa, Kutasov}.  There are also
purely field-theoretic constructions,
like the Intriligator-Seiberg-Shih (ISS) models \cite{ISS}
where strong/weak coupling dualities \cite{SeibergDuality}
allow one to find such states, or the retrofitted models \cite{Retro}
where canonical theories like O'Raifeartaigh or Polonyi models can be
coupled to additional dynamics
in a way that naturally produces an exponentially small SUSY breaking scale.
These states have potential applications
both in understanding the existence and properties of stable non-supersymmetric
string compactifications \cite{BP, Silverstein, KKLT} (for recent reviews see
\cite{DouglasKachru,Lust}), and in building realistic
models of gauge-mediation \cite{GM} or sequestered, high-scale SUSY
breaking \cite{Seq}.
It is natural to think that via AdS/CFT duality \cite{juanAdS} or brane engineering
\cite{Hanany-Witten}, one can sometimes relate the stringy and field
theoretic constructions.
Indeed, many groups have recently
engineered various D-brane field theories which exhibit
dynamical SUSY breaking (DSB) and reduce to
known DSB field theories in the decoupling limit
\cite{Ooguri:2006in,Franco:2006es,OoguriOokouchi, FrancoUranga, MQCDSeibergShih,
Argurio:2006ny, TatarMeta, Kitano:2006xg, Wijnholt, Antebi}.

Here, we continue our investigation \cite{Argurio:2006ny} into the possible relations
between ISS-like states in field theory, and SUSY-breaking states where SUSY
is broken at the end of a ``warped throat," as in \cite{KachruPearsonVerlinde} (where
SUSY was broken by anti-D3 brane probes at the tip of a warped, deformed conifold
\cite{KS}).  At a qualitative level, it is natural to think that SUSY breaking at
the end of a warped throat is AdS/CFT dual to dynamical supersymmetry breaking.
Then one should try to interpret the SUSY breaking states involving warped
antibranes, which can tunnel to supersymmetric states of the
same gravitational system \cite{KachruPearsonVerlinde}, as metastable states in the
dual SUSY field theory.  While it is not necessary that such a correspondence should
hold (since non-supersymmetric vacua are not protected upon extrapolation in the
't Hooft coupling $g_sN$), it would be interesting and suggestive to find examples
where such states can be argued to exist both in the
gravitational system and the field theory dual.

In \cite{Argurio:2006ny}, we argued that the gauge/gravity duals derived by studying
fractional branes in simple quotients of the conifold are a natural place to find such
a correspondence.  In a particular $\bZ_2$ orbifold of the conifold, we were able to
realize a close relative of SUSY QCD where the small mass parameter of the ISS models is
dynamically generated, and where the dual gravitational system plausibly admits
metastable anti-brane states.  However, that work left several open questions.  Firstly,
the SQCD model that could be realized had $N_f = N_c$, while it is in the free
magnetic range $N_c + 1 \leq N_f < {3\over 2}N_c$ that one is really confident of the
existence of the metastable states in the field theory.  Secondly, because we wish to
dynamically generate the small mass parameter of the ISS models, we must take care to
ensure that the dynamical masses are ${\it stabilized}$ against relaxation to
zero or infinity.  In the models of \cite{Argurio:2006ny}, this delicate question
rested entirely on (largely unknown) properties of the K\"ahler potential.

In this paper, we argue that a simple variant of the model of \cite{Argurio:2006ny} solves
both of these problems.  By considering other orbifolds of the conifold, we are able to
find models where we can reach the free magnetic regime necessary to prove
existence of the ISS vacua, and where we can argue that the superpotential itself helps
to stabilize the dynamical quark masses in the range where SUSY breaking occurs. More precisely,
the models we construct have $N_f=N_c+1$.
As an added bonus, our model shares a nice feature with the model of Kitano, Ooguri and
Ookouchi \cite{Kitano:2006xg}: the quiver superpotential naturally comes with terms that
break the R-symmetry of the original ISS models, which is problematic in model building
applications (as it forbids a gaugino mass).

In the rest of this section, we introduce our model. Its field theory dynamics is analyzed
in \S2, where we show how an effective massive SQCD arises in a given corner of its moduli
space. An important role is played by a stringy instanton
generated contribution to the
effective superpotential, whose origin we discuss in \S3. In \S4
we prove that the quark masses can be
dynamically stabilized, and in \S5
we estimate the lifetime of the metastable vacua. In \S6,
we briefly discuss the gravity dual IIB description (involving fluxes
and branes in the deformed geometry).  We also
present a IIA T-dual of the IIB picture, where the gauge theory arises from a
configuration of NS 5 branes and D4 branes.
These descriptions allow one to visualize many (though not all) aspects
of the field theory dynamics, and in particular, make it obvious that the metastable
SUSY-breaking vacua are dual to models which contain anti-branes. We conclude in \S7.

\subsection{The structure of the model}
\label{model}

The models we would like to analyze are obtained by considering (fractional) D3 branes
at the tip of a non-chiral $\bZ_N$ orbifold of the conifold, which is nothing but a
straightforward generalization of the system considered in \cite{Argurio:2006ny} for the case $N=2$.

The corresponding quiver gauge theory admits $2N$ gauge factors and $4N$ bifundamental chiral superfields $X_{ij}$
interacting via the following quartic superpotential
\be
\label{wtree}
W = h \sum_{i=1}^{2N} (-1)^{i+1} X_{i,i+1}X_{i+1,i+2}X_{i+2,i+1}X_{i+1,i}~,
\ee
where the index $i$ is understood $mod(2N)$.

Because the quiver is non-chiral, we can consistently assign arbitrary ranks to the quiver nodes, which
suggests that there should be $2N-1$ independent fractional branes one can define. This is indeed
mirrored in the geometric structure of the singularity, which admits $2N-1$ independent shrinking
2-cycles the branes can wrap. In a given basis, which will be relevant later, one obtains a natural classification
into $N$ deformation fractional branes and $N-1~$ ${\cal N}=2$ fractional branes, following the
definition proposed in \cite{seba}. We remind the reader that deformation
branes correspond to isolated nodes in the quiver (and hence gauge groups with no matter)
and lead to confinement, while ${\cal N}=2$
branes arise from occupying two connected nodes, which yields
a product of two SQCD theories with $N_f=N_c$, and hence have a moduli
space of vacua.

\begin{figure}[ht]
  \centering
  \includegraphics[width=4.5cm]{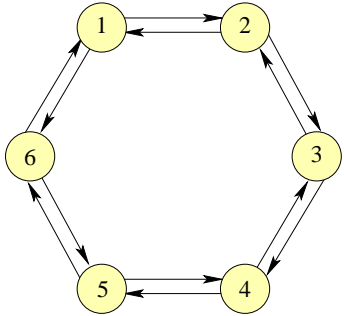}
  \caption{The quiver describing the field content of the gauge theory at the tip of the
  non-chiral $\bZ_3$ orbifold. The ranks of
the gauge groups can be chosen arbitrarily.}
  \label{quiverg}
\end{figure}

As we are going to show, for our present purposes it is enough to take $N=3$ (any
$\bZ_{N}$ with $N>3$ naturally works in the same way). Therefore,
from now on we stick to this case, for simplicity.
This specific $\bZ_{3}$ quotient admits five shrinking 2-cycles.  Two of them
see, locally, a ${\bC}^2/{\bZ}_2$ singularity.  The other three are dual (via geometric
transitions) to compact 3-cycles $A_i$ ($i=1,2,3$) which can be made finite by
a complex deformation.
The
corresponding quiver is shown in Figure
\ref{quiverg}.

\begin{figure}[ht]
  \centering
  \includegraphics[width=5.5cm]{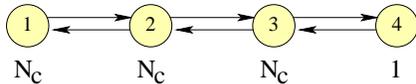}
  \caption{The quiver of the theory (\ref{5rank}).}
  \label{4nodes}
\end{figure}

We would like to consider the gauge theory arising from the following assignment of
ranks in the quiver
\be
\label{5rank}
(N_c~,~N_c~,~N_c~,~1,~0~,~0)~.
\ee
In terms of fractional branes, this may be viewed as $N_c$ ${\cal N}=2$ branes at nodes
1 and 2,  $N_c$ deformation
branes at node 3 and another single deformation brane at node 4 (this definition is basis-dependent, of course).
The``$SU(1)$" fourth node is not actually a gauge group. Its interpretation is that $X_{34}$ and $X_{43}$ transform purely as
fundamental and antifundamental representations of node 3, respectively.
The corresponding quiver is depicted in Figure \ref{4nodes}.

As we shall show, this system, whose dynamics we are going to study in detail, reduces exactly (in a region
of the moduli space to be specified below) to massive $SU(N_c)$ SQCD with $N_f=N_c+1$ massive (but light) flavors,
and therefore admits both the supersymmetric and the metastable non-supersymmetric vacua of that theory. Hence,
this system provides a string
embedding  of an ISS model. Moreover, it has some additional virtues: the small flavor masses are dynamically generated
(and stabilized), it
is possible to give a simple gravity dual interpretation of the metastable non-supersymmetric vacua,
and R-symmetry is explicitly broken (which could be useful in any model-building applications).

\section{The dynamics of the model}

The quiver gauge theory we are going to analyze is the one depicted
in Figure \ref{4nodes}. This theory has a superpotential of the form
\be
W = h( X_{12}X_{23}X_{32}X_{21} - X_{23}X_{34}X_{43}X_{32})
+ m X_{43} X_{34}~.
\label{wtree5}
\ee
As already discussed, the two quartic terms follow from the conifold by standard orbifold techniques.
The mass term for $X_{34}$ and $X_{43}$ is generated by a stringy instanton. We postpone
discussion of the relevant instanton to \S3, and
we proceed to analyze the above superpotential.
The quartic coupling $h$ has the dimensions of an inverse mass, and
is inversely proportional to the UV scale generating the non-renormalizable interaction.
In this context, it is natural to take $h \sim 1/M_s^*$. Here, $M_s^*$ indicates the string
mass scale effectively warped down to a lower value due to the RG flow,
which manifests itself as a duality cascade. For the field theory interpretation
to be valid, we need $M_s^*$ to be bigger than any of the dynamical scales of the gauge groups
involved in the quiver.

To start analyzing the gauge theory, we will make some assumptions about the
scales of the gauge groups on every node.
Node 3 is the main node where the ISS-like SQCD dynamics takes place.
Node 2 acts as a subgroup of the flavor symmetry, broken
as $SU(N_c+1)\times SU(N_c+1) \supset SU(N_c+1) \supset SU(N_c)$.
Accordingly, we will take its dynamical scale $\Lambda_2$ to be (much) smaller
than the others, so that this gauge group
can be effectively considered as classical.

Node 1, which has a number of flavors which equals the number of
colors, undergoes confinement so that its effective dynamics is described
in terms of mesons and baryons. The mesons are going to supply mass terms
for some of the flavors of node 3, through the superpotential couplings
in (\ref{wtree5}), much as in \cite{Argurio:2006ny}. These masses are
subject to the constraint on the deformed moduli space of node 1.
In particular, we will need the scale $\Lambda_1$ to be such
that those masses are still lower than the scale of the SQCD node, $\Lambda_3$.
The additional mass $m$ will be larger than these $N_c$ masses, but still
smaller than $\Lambda_3$. We will explain later how this parameter
range can be obtained.

This model is quite similar to the one in
\cite{Kitano:2006xg}.  One difference is that all the small parameters are generated
dynamically. In addition, this model arises naturally at a fairly simple Calabi-Yau
singularity.

Assuming now that node 1 is confining, the tree level superpotential
reads
\be
\label{wlec}
W = h ( M_{22} X_{23}X_{32}  - X_{23}X_{34}X_{43}X_{32})
+m X_{43}X_{34}~,
\ee
where we have defined $M_{22}=X_{21}X_{12}$. This superpotential is not quite complete:
we should really implement the constraint relating the mesons and the
baryons of node 1 through the introduction of a
Lagrange multiplier. We delay that
to later on. Let us now assume that the interactions are such that
the mesonic and baryonic branches of node 1 decouple, and in particular that
when the meson matrix has maximal rank the baryons are required to vanish --
we will argue that this is the case in \S\ref{stabilization}.
For the time being, we assume that node
1 is on the mesonic branch, where the constraint
describing the quantum-deformed moduli space reads
\be
\det M_{22} = \Lambda_1^{2N_c}~.
\label{wewant}
\ee
This constraint is necessary for the generation of dynamical masses but does not
fully fix the eigenvalues of
$M_{22}$. In the non-supersymmetric vacua, their stabilization occurs at tree-level as we explain shortly.
At the stable point, the VEV of $M_{22}$ is proportional to
the identity matrix. Hence we see that in the effective SQCD theory at node 3,
we have $N_c$ flavors of mass $\sim h \Lambda_1^2$
and $1$ flavor of mass $m$. We will take the latter
to be the heavier one, so that $h\Lambda_1^2 < m$.

Therefore, along this branch the theory on node 3 with superpotential (\ref{wlec}) is
nothing but SQCD with $N_f = N_c+1 $ massive flavors, together with a quartic coupling
(which is irrelevant in the IR).
Integrating out the flavors, we obtain pure $SU(N_c)$ SYM characterized
by a dynamical scale
\be
\Lambda_L^{3N_c} = \Lambda_3^{2N_c-1}h^{N_c} \det M_{22} ~m
= \Lambda_3^{2N_c-1}(h \Lambda_1^2)^{N_c} m~.
\ee
Implementing the constraint on the deformed moduli space of node 1
with a Lagrange multiplier in an effective superpotential, it is easy to
see that we indeed have a moduli space of supersymmetric vacua where
$M_{22}$ has non zero VEV, while the baryons
are vanishing. When taking into account that node 2 is actually
gauged, we see that at low-energies the moduli space will be described
by $\mathbf{C}^{N_c-1}$ together with a residual $U(1)^{N_c-1}$ gauge symmetry.

We now move on and show that our theory also has meta-stable, SUSY breaking vacua.
Since node 3 has $N_f=N_c+1$, its low-energy dynamics is governed by a theory
of mesons and baryons. This case can actually be seen as a limiting
case of Seiberg duality, where the dual magnetic gauge group is a trivial
$SU(1)$, and the dual quarks are nothing but the baryons of the electric theory.
In the following, we will adopt this terminology.
The bifundamentals are combined
into effective mesons as $X_{i3}X_{3j}= \Lambda_3 \phi_{ij}$, and the
dual quarks are labeled $Y_{i3}$ and $Y_{3i}$. The superpotential is
\be
W=(h\Lambda_3) (M_{22}\phi_{22} - \Lambda_3
\phi_{24}\phi_{42}) + m\Lambda_3  \phi_{44}
-\phi_{22}Y_{23}Y_{32} - \phi_{44}Y_{43}Y_{34}  +\phi_{24}Y_{43}Y_{32}
+\phi_{42}Y_{23}Y_{34}~.
\label{wmag}
\ee
Note that we have rescaled the mesons to canonical dimension using the scale
$\Lambda_3$. Accordingly, the cubic terms generated by the duality
have a coupling of $\cO(1)$, which we set to one (shifting the undetermined
constant to the normalization of the canonical K\"ahler potential).
Strictly speaking, we should also add a term linear in the determinant
of the meson matrix, but it is highly irrelevant in the IR (and to our
considerations).\footnote{It does play a role if we want to recover the
SUSY vacua in the low-energy picture.} It is easy 
to see that the above model of mesons and dual quarks would
have, in the absence of the $\phi_{24} \phi_{42}$ coupling, an accidental IR
$U(1)_R$ symmetry. The R charges would be 2 for the
mesons and 0 for the dual quarks, as in an O'Raifeartaigh model.
The quadratic coupling in the mesons which
arises naturally in the present model provides an explicit
breaking of this R-symmetry.

This theory is now amenable to an analysis very similar to
\cite{ISS,Kitano:2006xg}. There is supersymmetry breaking by the rank
condition. The F auxiliary field that vanishes is the one related to the
more massive flavor,
i.e. the F-term of $\phi_{44}$. On the other
hand, the F-components of $\phi_{22}$ are non-vanishing. As a consequence,
there is a tree level vacuum energy given by
\be
V_{tree} = |h\Lambda_3|^2 \sum_{i=1}^{N_c} | M_i |^2 = N_c |h\Lambda_3
\Lambda_1^2 |^2~,
\ee
where we obtain the final result by extremizing on the eigenvalues of the
matrix $M_{22}$ given the constraint on its determinant.
We are going to show later that indeed the constraint on the determinant
is not destabilized by baryonic VEVs.

A standard analysis of this model shows that it is
a sum of O'Raifeartaigh models with an additional coupling $m_{24}=h\Lambda_3^2$, which
is the one quadratic in the two mesons $\phi_{24}$ and $\phi_{42}$.
As we will discuss in \S\ref{stabilization}, $\phi_{22}$ gets a non-zero VEV
due to the one-loop potential (see also \cite{Kitano:2006xg}).
This VEV is directly related to the presence of
the quadratic meson coupling $m_{24}$, so that we can actually estimate it as
\be
|\phi_{22}| \sim |h \Lambda_3^2|~.
\ee
As noted in \cite{Kitano:2006xg}, the vacua analyzed here are unstable
if the size of $m_{24}$ exceeds the larger flavor mass.
Here this bound reads
$|h \Lambda_3^2|^2 < |m\Lambda_3|$, or in other words
(recall that $h=1/M_s^*$)
\be
\left(\frac{\Lambda_3}{M_s^*}\right)^2 < \frac{m}{\Lambda_3}~.
\label{mzvsm}
\ee
Note that all these relation must be taken with a grain of salt, since
there are factors of $\cO(1)$ that we are not retaining (most
of which are non-calculable anyway).

All other pseudomoduli are lifted by the
one-loop potential, and acquire a non-tachyonic mass.

This shows that the present model has metastable supersymmetry
breaking vacua, provided we can show that there is no instability
towards turning on baryonic VEVs at node 1.
We demonstrate in \S\ref{stabilization} that this is the case, in an appropriate regime
of
parameters.\footnote{In the model that we presented
in \cite{Argurio:2006ny} there is potentially such an instability.  In that model
the SQCD node is in a confining rather than IR free regime, so that
(non-calculable) corrections to the K\"ahler potential are present,
and make it difficult to determine what happens.
The gravity dual description provides another source of information; such
an instability is not readily apparent there, but it is a complicated
system which would benefit from further study.}

\section{A mass term generated by a stringy instanton}
\label{instanton}
Before analyzing in detail the stability of the model, we have to
explain how the mass term for the additional flavor is generated.

It turns out that
somewhat novel stringy instanton effects which have recently been
investigated in several other contexts \cite{FKMS,Cvetic,UrangaIbanez,
Haack,Bianchi:2007fx,cvetic,abflp}
contribute corrections to $W$ which depend on gauge invariants  that
usually do not appear in the superconformal quiver superpotential.
Recall that these effects arise when
Euclidean D branes wrap cycles corresponding to quiver nodes which are
{\it not} occupied by space-filling branes. In this respect, they
are specific to set ups with fractional branes. We now show that
such an instanton generates the mass term $m$.

\begin{figure}[ht]
  \centering
  \includegraphics[width=6cm]{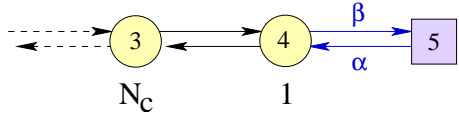}
  \caption{The extended quiver describing the
  interaction between the fractional brane system and a Euclidean
D1 brane on node 5, represented by a square. The relevant coupling is a quartic one, involving $\alpha,\beta, X_{34}$
and $X_{43}$, \eref{instac}.}
  \label{inst}
\end{figure}

To understand the instanton contributions,
consider a D1 instanton (a Euclidean D1 or ED1 brane)
wrapping node 5 of the quiver.
It is BPS and preserves precisely 1/2 of the ${\cal N}=1$ supersymmetry;
acting on the instanton solution with the broken supercharges then produces
two fermion zero modes in the ED1 - ED1 open string sector.  These are
the two fermion zero modes that are necessary to give rise to a contribution
to the space-time superpotential (we discuss the possibility of extra
``accidental'' zero modes at the end of this section).
Considering the extended quiver diagram including a node for the instanton, there are also fermionic
strings $\alpha, \beta$ stretching to node 4, in the $(+,\overline{N_4})$ and
$(-,N_4)$ representations of the $U(1) \times SU(N_4)$ gauge group (in the
present case we will have $N_4=1$). As follows from
the computations in \cite{FKMS,UrangaIbanez}, the fermionic spectrum in the extended quiver
is the same as it would be if
the instanton were actually a space-filling brane, except the fermions live in a different
dimension. By the simple argument of \cite{UrangaIbanez}, we also expect that there are
no bosonic zero modes in this sector. Bosons would arise from NS sector strings,
but the NS sector ground state energy receives a contribution from the number of
ND boundary conditions, which pushes the ground state energy above zero
in this configuration. The relevant part of the
extended quiver is reported in Figure \ref{inst}.

In the instanton action, we expect a gauge invariant coupling
\begin{equation}
\label{instac}
L = \alpha X_{43} X_{34} \beta~.
\end{equation}
In evaluating the instanton contribution to the 4d effective action, we should
integrate over the only two charged fermionic zero modes $\alpha, \beta$.
This yields a simple contribution to the superpotential
\be
\label{wmass}
c ~X_{43} X_{34} ~e^{-{\rm Area}}.
\ee
where $c$ is a dimensionful constant and the relevant
area is the area of the curve corresponding to node 5.
We thus identify the mass term as $m=c ~e^{-{\rm Area}}$. If we reasonably
take $c\sim M_s^*$, we see that it is not difficult to assume that
the area of the cycle wrapped by the instanton is such that $m<\Lambda_3$.
(Roughly that would amount to assume that $\Lambda_5\ll \Lambda_3$ if there was
a gauge group on node 5.)

Now, a similar instanton on node 6, with fermionic strings stretching
to node 1, produces another term in the superpotential.
The gauge invariant coupling in the instanton action is
\begin{equation}
\label{instac6}
L = \alpha X_{12} X_{21} \beta~.
\end{equation}
If we let $a,b$ denote gauge indices at node 1
and $f,g$ denote gauge indices at node 2, then the gauge contractions in
(\ref{instac}) give rise to $\alpha_{\bar a}X_{12}^{a\bar f} X_{21, f\bar b}
\beta^{b}$.  So performing the integral over the $\alpha, \beta$ fermions,
which are now a set of $2N_c$ zero modes,
gives the contribution
\begin{equation}
\label{barmass}
c' B \tilde B ~e^{-{\rm Area}'}
\end{equation}
to the 4d effective theory, where $c'$ is similarly a dimensionful constant
and the area is now the one of the curve corresponding to node 6.
The coupling (\ref{barmass}) provides
a mass term for the baryons $B=\det X_{12}$ and $\tilde B=\det X_{21}$ of node 1.
We will see in the next section that this term does not however play
an important role in the stabilization of the baryons.

Let us end this section with a comment regarding a subtle point. With the above reasoning, the coefficients
$c$ and $c'$ have been determined up to a dimensionless number whose precise value
we cannot directly compute
in our geometric set-up. A crucial
ingredient for such coefficients not to
vanish involves
the number of uncharged fermionic zero modes on the ED1 brane.
Before taking into account the quiver branes back-reaction, there
are four, since
the ED1 is a 1/2 BPS state in the Calabi-Yau. While, as already discussed, two zero modes
are necessary to provide the chiral superspace integral for the superpotential term
(\ref{wmass}), the other two would provide a dangerous vanishing contribution.
Still, one should take into account the full back-reacted closed string
background, which includes non-trivial fluxes. This background preserves only 4
supercharges out of the 8 preserved by
the CY, so that the instanton has only 2 zero modes associated to
broken supersymmetries. Then, at least in many backgrounds,
it is reasonable to expect that the extra
zero modes get lifted by the interactions with other background fields.
This is an interesting problem in itself, but we leave it
for further research.
Instead, in order to provide a background where we can
explicitly identify an
object responsible for lifting the additional zero modes,
we can introduce orientifold planes in such a way that the instanton
wraps a cycle that is mapped to itself. In this way, half of the zero modes
are projected out from the start.  One concrete
embedding of our model in an orientifold that accomplishes this
task, while not spoiling all other nice features of our model, is described in
Appendix \ref{appendix_orientifold}.
We therefore conclude
that the coefficient $c$ is non-vanishing in many suitable models.\footnote{
A more complete discussion of these issues, for backgrounds
where a simpler worldsheet CFT description is available, will appear in
\cite{abflp}. A general discussion on the introduction of O-planes
in generic Calabi-Yau geometries will appear in \cite{francoetal}.}

\section{Stabilizing dynamical masses}

\label{stabilization}
We have explained how our theory has metastable supersymmetry breaking vacua
under certain assumptions regarding the stability of the dynamically generated
masses. An important question is whether the dynamical masses relax to zero by turning on expectation values
for the baryonic fields at node 1. We show in this section that these
a priori dangerous directions are lifted because
$\langle \phi_{22} \rangle\neq 0$. New superpotential interactions
generated by the D1 instanton wrapping node 6 of the quiver also contribute to
stabilization, although they are not the dominant effect.

We thus first sketch how the one-loop potential gives
the crucial VEV to the field $\phi_{22}$. We start from (\ref{wmag}) and expand around the metastable vacuum. This is
characterized by VEVs for $Y_{34}, Y_{43}=\sqrt{m\Lambda_3}$ and, at tree
level, by an arbitrary $\phi_{22}$. The latter is also the field with
non-vanishing F-terms. The superpotential
for the fluctuations of the fields, expanded about the vacuum,
takes the form
\be
W = m_{22}^2 \phi_{22}  - Y_{32} \phi_{22} Y_{23}
+ m_{44} \phi_{24} Y_{32} + m_{44} \phi_{42} Y_{23} - m_{24} \phi_{24}\phi_{42}~,
\ee
In writing this down, we have dropped several fields: $\phi_{44}$,
$\delta Y_{34}$
and $\delta Y_{43}$ do not feel SUSY breaking at this order so will cancel out
of the one-loop energy. The masses appearing above are given by
$m_{22}^2= h\Lambda_3 \Lambda_1^2$, $m_{44}^2= m \Lambda_3$ and $m_{24}=
h \Lambda_3^2$. All fields have a canonical K\"ahler potential.

It is straightforward to realize that the F-terms
set $\phi_{22}$ to a diagonal form. Then, we obtain
just a superpotential for $N_c$ decoupled O'Raifeartaigh
like models. Each such model has, besides the field with the non-zero
F-term, 4 other fields.
An important parameter is the coupling $m_{24}$ of the meson bilinear
$\phi_{24}\phi_{42}$.

The tree level vacuum energy is just:
\be
E_{vac} = N_c \,|m_{22}|^4~.
\label{E_vac_0}
\eeq
We had already noted that the eigenvalues of $M_{22}$ are all trivially
stabilized at their common values \cite{Argurio:2006ny}.

To compute the one-loop vacuum energy of this model, we simply compute the boson/fermion masses as a
function of pseudo-moduli using
\beq
m_0^2=\left(\begin{array}{ccc} W^{\dagger ac}W_{cb} & \ & W^{\dagger abc}W_c \\
W_{abc}W^{\dagger c} & & W_{ac}W^{\dagger cb} \end{array} \right) \ \ \ , \ \ \
m_{1/2}^2=\left(\begin{array}{ccc} W^{\dagger ac}W_{cb}& \ & 0 \\ 0 & & W_{ac}W^{\dagger
cb} \end{array} \right)
\label{m_scalars_fermions}
\eeq
and plug them into the famous 1-loop Coleman-Weinberg result.
The eigenvalues of both the fermionic and the bosonic mass square matrices
can be computed analytically. In this model $\phi_{22}$ remains massless
at tree level and its Fermi partners do too. The other 8 eigenvalues
split in pairs as follows. The bosonic ones are given by
\be
|m_{44}|^2 +\frac{1}{2}(|\phi_{22}|^2+|m_{24}|^2\pm |m_{22}|^2) \pm'
\frac{1}{2}\sqrt{(|\phi_{22}|^2-|m_{24}|^2\pm |m_{22}|^2)^2 +
4|m_{44}\phi_{22}^\dagger + m_{44}^\dagger m_{24}|^2}~,
\ee
with the $\pm$ and $\pm'$ standing for two independent sign choices.
The fermionic ones are given by the same expression as the bosonic
ones, save that we formally set $m_{22}=0$ (the fermionic sector does not
talk directly to the F-term).

If we set $m_{24}=0$ we obtain the classic O'Raifeartaigh result
for the one-loop energy, i.e.
\beq
E_{1} = {N_c \over 32\pi^2}|m_{22}|^4 \left( y^{-2} (1+y)^2 {\rm log}(1+y) +
y^{-2} (1-y)^2 {\rm log} (1-y) + 2 {\rm log} ({|m_{44}|^2\over \Lambda^2})
\right)~,
\label{E1_zeroth_order}
\eeq
where we have defined $y = \vert{m_{22} \over m_{44}}\vert^2$ and $\Lambda$ is
the UV cutoff. In this case, $\phi_{22}$ stabilizes around zero.

When $m_{24}\neq 0$, the analytical form of the one-loop energy is not
very illuminating. However it is reasonable to expect that $\phi_{22}$
will pick up a tadpole around zero, so that its vacuum expectation
value is displaced to a non-zero value. Moreover, the size of the
VEV is directly
controlled by $m_{24}$, as they enter almost symmetrically in the
expressions for the eigenvalues. This is confirmed by a numerical analysis,
which also shows the existence of tachyons when $m_{24}$ is too close
to or larger than $m_{44}$. Indeed, one might have guessed that in this
range some dangerous mixings can occur.

Let us remark at this stage on a possibility which could have been
considered. We could have actually tried to generate the higher masses
like $m_{44}$ dynamically in the same way as the lower ones, $m_{22}$.
That would be simply implemented in a 5-node quiver with
ranks $N_f-N_c$ at the 4th and 5th node. The model would be very similar
to the above, except that every O'Raifeartaigh model would have now
$1+4(N_f-N_c)$ fields. If the masses were dynamical, one would have
a sum of $N_f-N_c$ contributions like (the generalization to $m_{24}\neq 0$ of)
eq. (\ref{E1_zeroth_order}). The latter potential attracts all
of the higher masses to smaller values. However one can see that the
deformed moduli space constraint is not sufficient
in this case to stabilize them. Indeed,
in the dominant contribution (the $\log |m_{44}|^2$ piece), the constraint
gives trivially a constant. The rest of the potential asymptotes  to
a constant for very large $m_{44}$. Hence, it will always be favorable
to bring down some masses while sending the other(s) to infinity.
This is the reason why we cannot really access the full IR free region
$N_c+1 \leq N_f < {3\over 2} N_c$ of SQCD in this class of models.

\subsection{Stabilization of baryonic directions}

\label{section_stabilizing_masses}

At this stage, we have seen that as long as we can be on the mesonic branch
at node 1, we will successfully obtain a model of metastable
supersymmetry breaking with no small parameters added by hand.

An important question now arises, however.
Assuming we are working in the regime $h\Lambda_1^2 < m$, the energy of
the SUSY breaking vacuum is given by
\begin{equation}
V \sim N_{c}\,h^2 \Lambda_{3}^2 \,\Lambda_{1}^4 ~,
\label{vmeta}
\end{equation}
since it arises by summing the masses squared of the $N_c$ lightest flavors
of the $SU(N_c)$ gauge group at node 3 \cite{ISS}.
The quantum moduli space constraint for the mesons $M_{22}$ is really
\begin{equation}
\det M_{22} - B \tilde B = \Lambda_{1}^{2N_{c}}~.
\end{equation}
So, at least naively, it appears that by relaxing the mesonic VEVs and turning on
$B, \tilde B$, one can lower the vacuum energy to zero, destabilizing the
SUSY breaking vacuum.  It is conceivable that the K\"ahler potential (which is
not computable) introduces a barrier that prevents such relaxation, but confidence
in the construction would be considerably enhanced if additional superpotential
terms were present which prevent
the baryons from `turning on' when one expands around the point (\ref{wewant}).

We now begin estimating the leading $B, \tilde B$ mass matrix by expanding
the potential around the would-be non-supersymmetric vacuum. To do so, we assume a canonical K\"ahler potential. We consider this is reasonable, since both potential instabilities and stabilizing effects arise under this assumption.
The leading off-diagonal term is
\beq
V_{,B \tilde{B}}=V_{,\tilde{B}B}=- \, h^2 \, \Lambda_3^2/\Lambda_1^{2N_c-4}~.
\label{off_diagonal_terms}
\eeq
This contribution appears at tree-level and favors the condensation of baryons
as discussed above.

However, there are several further terms in the (super)potential which
impart diagonal terms in the mass matrix, and overwhelm the tachyonic
contribution (\ref{off_diagonal_terms}) for reasonable choices of parameters.
One source of such a term is the tree-level coupling of $\phi_{22}$ in
(\ref{wmag}).  This will in fact turn out to be the dominant effect
stabilizing $B, \tilde B$ at zero.
For completeness, we also include the sub-dominant effect caused by
the stringy instantons discussed in the previous section.

Putting (\ref{barmass}) together with (\ref{wmag}), we can check for stability of the mesonic
branch VEVs (\ref{wewant}) as follows.  Assume the $M_{22}$ matrix is diagonal,
with equal eigenvalues given by $x$.  $x$ is then determined by the deformed quantum
moduli space constraint of the node 1 gauge theory to be
\begin{equation}
x^{2N_c} = \Lambda_{1}^{2N_c} - B \tilde B~.
\end{equation}
We could impose this constraint by adding a Lagrange multiplier $\lambda_1$ to the
superpotential, multiplying the constraint equation.  Then, subject to the constraint,
we should minimize the potential
\begin{equation}
\label{potis}
V = \Lambda_1^2 \, \vert h \, \Lambda_3 \, \phi_{22} + \lambda_{1} x^{N_c-1}\vert^2 + \Lambda_1^{2N_c-2}\,
\vert \lambda_1 B + c_1 B\vert^2 + \Lambda_1^{2N_c-2}\, \vert \lambda_1 \tilde B + c_1 \tilde B\vert^2~,
\end{equation}
where the first term arises from $|F_x|^2$, the second term from $|F_{\tilde B}|^2$,
and the third from $|F_{B}|^2$, and we have redefined $c_1 = c'~e^{-{\rm
Area}'}$ with respect to (\ref{barmass}).

Equation \eref{potis} only contributes to the diagonal
\begin{equation}
\label{baryonmass}
V_{,BB} = V_{,\tilde B \tilde B} = {2\over \Lambda_1^{2N_c+2}} ( c_1 \Lambda_{1}^{2N_c} -
h \Lambda_1^2 \Lambda_3 \phi_{22})^2 ~.
\end{equation}

From \eref{baryonmass} we can obtain the leading diagonal terms in the matrix of second derivatives of the potential. The leading
contribution is a non-zero expectation value $\phi_{22}\sim m_{24}=h \Lambda_3^2$. The net result is

\beq
V_{,BB}=V_{,\tilde{B}\tilde{B}}=h^4\Lambda_3^6/\Lambda_1^{2N_c-2}.
\label{diagonal_terms}
\eeq

Both $h$ and the instanton coefficient $c_1$ are suppressed by $M_s^*$, as $h\sim M_s^{*-1}$ and
$c_1\sim M_s^{*3-2N_c}$. (There is also a suppression by the volume of the
curve representing node 6 for the instanton, but since this effect plays
no role in our theory as the $\phi_{22}$ VEV already stabilizes the baryons of
node 1, we can take that volume to be anything
$\geq {\cal O}(1)$. For simplicity we've chosen ${\cal O}(1)$ here).
The only consistency requirement on the relevant scales is then
that $M_s^* > \Lambda_3$.

The eigenvalues of the matrix of second derivatives of the potential are $V_{BB}\pm V_{B\tilde{B}}$.
Then, we are free of tachyons provided that $V_{BB}\geq |V_{B\tilde{B}}|$. From \eref{off_diagonal_terms}
and \eref{diagonal_terms} we conclude the conditions for stability of the baryonic directions are
\beq
(\Lambda_1/\Lambda_3) < (\Lambda_3/M_s^*)
\label{stability_baryons}
\eeq

We see that we can always satisfy the above inequality, as well
as the ones coming from the hierarchy of mass scales in the low-energy model
(as in e.g. \cite{Kitano:2006xg}), by imposing the following hierarchy
\be
\Lambda_1 \ll  \Lambda_3 < M_s^* ~~\mbox{and}~~ m < \Lambda_3~.
\ee
As discussed previously, we also need $m$ to satisfy the bound \eref{mzvsm}.

To summarize, we have checked that the potential baryonic instability
is cured. To do this, the one-loop generated VEV of $\phi_{22}$
is enough. On the other hand, to generate $m$, a crucial role was played by
an additional term in the superpotential, generated by a string instanton.
This makes our model a bona-fide version of SQCD with dynamically
generated exponentially small quark masses, and allows it to stably
display the related metastable vacua.

\section{Lifetime of the meta-stable vacuum}

In this section we study the possible decay channels for our non-supersymmetric ISS-like vacuum.

As reviewed in \S2, there is a SUSY vacuum where the mesons of node 3
acquire VEVs, which in turn are fixed by the VEVs of the mesons of node 1.
This is what we refer to as the mesonic branch, and is the usual SUSY
vacuum of SQCD, as considered e.g. in \cite{ISS}.
In addition to this SUSY vacuum, there is also another direction of possible
decay, which is precisely the one discussed in the previous section. Along
this direction, the baryons of node 1 acquire VEVs, and we are essentially
led to an SQCD at node 3 with $N_c$ massless and one massive flavors.
By standard arguments used when discussing
the deep IR of cascading quivers, one can see that
after a Seiberg duality on node 3,
the quiver reduces to SQCD with one flavor at node 2. The flavor acquires
a mass which is directly related to $m$.

We thus want to estimate the decay rate towards these two (classes of) vacua.
We do this by estimating the bounce action in the following form,
using the triangular approximation \cite{dj}
\be
S \sim \frac{(\Delta \Phi)^4}{\Delta V},
\ee
where $\Delta \Phi$ is roughly the width of the barrier while
$\Delta V$ is its height. We will see below that
we do not need to be more precise, since we are really interested
in lower bounds for the bounce action anyway. If we can tune these
lower bounds to be large enough, we can be confident that decay through
tunneling is suppressed and meta-stability is not affected.

Let us first consider decay towards the mesonic branch SUSY vacua.
Here $\Delta V$ is readily evaluated to be of the order of \eref{vmeta}, since
the energy of the metastable state and the peak only differ by a numerical
factor. To estimate $\Delta \Phi$, we first note that, as in \cite{ISS},
the fields which have the biggest variation are the mesons of node 3.
In the SUSY vacuum, their VEVs are given by
\be
\phi_{22} = \Lambda_3 \left(\frac{m}{\Lambda_3}\right)^\frac{1}{N_c}, \qquad
\qquad \phi_{44} = \Lambda_3 \frac{h\Lambda_1^2}{m}
\left(\frac{m}{\Lambda_3}\right)^\frac{1}{N_c}.
\ee
It is obvious from the above that $\phi_{22}\gg \phi_{44}$ for the
range of parameters discussed in the previous sections.
Moreover, recall that in the metastable state $\phi_{22}$ had a VEV
of the order of $h\Lambda_3^2$. This is however very small with respect
to the VEV it has in the SUSY vacuum, since we assume that $h\Lambda_3=
(\Lambda_3/M_s^*) \ll 1$. We can then identify $\Delta \Phi$
with the VEV of $\phi_{22}$ listed above.

Putting all together, we have the following estimate for
the bounce action towards the mesonic branch
\be
S_\mathrm{mesonic} \sim \left(\frac{m}{\Lambda_3}\right)^\frac{4}{N_c}
\left(\frac{\Lambda_3}{\Lambda_1}\right)^4
\left(\frac{M_s^*}{\Lambda_3}\right)^2.
\ee
Every factor in the expression above is (much) greater than one,
and we thus conclude that decay towards the mesonic branch is highly
suppressed.

As for the decay towards the baryonic branch of node 1, let us
provide the most conservative estimate. The field which varies the most
along the path is taken to be a representative baryon $B$. Its variation,
after the field has been canonically normalized, is taken to be
$\Delta \Phi \sim \Lambda_1$. Note that since $\Lambda_1$ is the smallest
scale in the game, this is really the most adverse situation.
As for $\Delta V$, we can take \eref{diagonal_terms} and plug in
the maximal VEV of the baryons $B \sim \Lambda_1^{N_c}$, so that
we get $\Delta V \sim h^4 \Lambda_3^6 \Lambda_1^2$. (Note that this
$\Delta V$ is much larger than the energy of the metastable vacuum.)
The bounce action is thus
\be
S_\mathrm{baryonic} \sim \left(\frac{\Lambda_1}{\Lambda_3}\right)^2
\left(\frac{M_s^*}{\Lambda_3}\right)^4.
\ee
Consistently with the inequality \eref{stability_baryons}, the above
bounce action can be made parametrically large, and thus also
the decay towards the baryonic branch is suppressed.

In this crude estimate, it seems that the latter
decay channel is the dominant one. To conclude that this is really
so would require a more serious investigation
of the potential and tunneling path. In any case we see that the
simplest estimates indicate that the apparent decay channels are
parametrically suppressed.

\section{The string dual description}

As we have seen, the quiver gauge theory we have analyzed in previous
sections admits a number of supersymmetric as well as metastable non-supersymmetric
vacua. A natural question is whether is it possible to provide a supergravity/string
dual description of such vacua. The well defined type IIB string embedding of
our model outlined in
\S \ref{model} makes this a realistic task.

\subsection{On the gravity dual}
\label{sugra}

In what follows we sketch the structure of only those vacua which are most relevant to our story:
the ISS-like vacua and their supersymmetric counterparts, i.e. the vacua
belonging to the mesonic branch. The discussion is very similar to the
one for the $\bZ_2$ conifold quotient presented in our previous work \cite{Argurio:2006ny}, to which we refer
for details.

In the present construction a crucial role is
played by the presence of an additional fractional brane, that we have to
treat as a probe since its backreaction cannot be captured classically
(by definition, we cannot take the large $N$ limit for a single brane).
Moreover, a second equally crucial ingredient is that there is a
mass term constraining the position of this probe brane. Indeed, it can
be checked that for $m=0$ both classes of supersymmetric vacua become
runaway. Again, the mass term is the product of a stringy instanton
which is not expected to backreact on the classical geometry in any
simple manner.

Below, we will take the pragmatic point of view that, because of the
mass term, we can roughly integrate out the effect of the additional
probe brane. We are then left with the same gravity dual as discussed
in \cite{Argurio:2006ny}, albeit embedded in a higher singularity.
The effects of the additional fractional brane presumably show up as
(important) $1/N$ corrections to the geometry.

Our brane system can be embedded into a weakly curved gravity dual
background by adding (a large number of) regular D3 branes. One can easily 
show that the resulting fractional/regular
brane system enjoys a duality cascade, i.e. a non-trivial RG-flow along which the effective
number of regular branes diminishes (in units of $N_c$, in this case). Hence,
choosing $N= k N_c$ regular D3 branes (with $k$ as large as we
wish), the IR field theory at the end of the cascade will be the quiver
field theory we have been studying.  The number of cascade steps $k$ will
just control the final warp factor in the IR region of the gravity dual.

As discussed in \S2, the region of the moduli space where an effective massive
SQCD $SU(N_c)$ theory emerges is along the mesonic branch of node 1.
This corresponds to having the $N_c$ ${\cal N}=2$ branes at a distance
$(\det M_{22})^{1/N_c} \propto \Lambda_1^2$ along $\bC$, the complex direction parametrizing the
VEV's of the adjoint scalar of the corresponding effective $SU(N_c)$ ${\cal N}=2$
vector superfield.
We will refer to the ${\cal N}=2$ fractional
branes as wrapped D5's, throughout this section.

The $\bZ_3$ orbifold of the conifold\footnote{As already mentioned, we can
embed the same 4-node quiver in a $\bZ_N$ orbifold with $N>3$. A reason to
go to larger $N$ might be to achieve the desired range of
scales, since there would be more geometrical quantities to tune.}
is described by the following equation in $\bC^4$
\be
\label{undefgeo}
x^3 y^3 = uv~.
\ee
As already discussed in \S \ref{model}, this geometry supports three independent complex deformations, leading to
the completely smooth geometry
\be
\prod_{i=1}^3 (xy-\epsilon_i) = uv~.
\ee
Consequently, there are three non trivial 3-cycles $A_i$ whose minimal
size is given by $\epsilon_i$
\be
\int_{A_i} \Omega = \epsilon_i~.
\ee
In our particular case we would like to consider the case where
only two of the three 3-cycles are blown-up, and moreover they have the same size \cite{Argurio:2006ny}
\be
(xy-\epsilon)^2 xy= uv~.
\ee
This deformation is triggered by the $N_c$ deformation branes we have at node 3.
The geometry above has a $\bC^2/\bZ_2$ line of singularities
(also called $A_1$-singularities, not to be confused with the label
of the 3-cycles above) at the locus $xy=\epsilon$, $u=v=0$. Moreover,
it has an innocuous conifold singularity at
$x=y=u=v=0$.

We construct the geometrical dual to the supersymmetric vacua of our theory,
which were discussed at the beginning of \S2, in the following way.
After a geometric transition, we expect the brane at node 3 to transmute and
turn into flux,
\be
\label{susyflux}
\int_{A} G_3 = N_c~~,~~ \int_{B} G_3 = \frac i g_s k
\ee
where $A$ is the compact 3-cycle absorbing the RR flux of the original branes,
$B$ its non-compact dual, and $k$ is the
number of duality cascade steps (and can be naturally taken to be very
large).
As already noticed, the D5's wrapping the $\bC^2/\bZ_2$ singularity
are instead explicitly present in the dual geometry, lying somewhere
along the mesonic branch. Finally, the
single deformation brane at node 4 cannot transmute,
and remains as a probe at the remaining conical singularity,
which we expect to be slightly deformed
by the instanton discussed in \S\ref{instanton}.
It can be checked that the above set up has the
same charges and supersymmetric moduli space as our theory.

We now move to the description of the metastable state. As originally discussed in \cite{KachruPearsonVerlinde}, and
recently applied in similar contexts by many authors, a natural way to construct metastable non-supersymmetric vacua
is by adding anti-branes. The positive vacuum energy is proportional to the number of such branes.
The fact that the vacuum energy is exponentially small is due, in the gravity dual, to the
warping of the anti-brane tension.

In order for these configurations to describe states in the same gauge theory,
one should check that the supergravity charges, at
infinity, are unchanged. In the present context this can be achieved by adding $N_c$ anti-D3 branes and simultaneously
jumping the NS fluxes by one unit
\be
\label{nsusyflux}
\int_{B} G_3 = \frac i g_s k \longrightarrow \int_{B} G_3 = \frac i g_s (k+1)~,
\ee
so as to leave the full D3 brane charge untouched.\footnote{Let us remind the reader that the full D3 charge is
$Q_3 = \int H_3 \wedge F_3 + N - \overline{N}$, where $N$ and $\overline{N}$ are the net number
of D3 and anti-D3 branes, respectively.}
It is a nice check of our proposal that it is only by adding $N_c$
such branes (no more, no less), hence providing the correct energetics
for the ISS metastable vacua, that we can leave the global charges
at infinity untouched, and hence describe non-supersymmetric
states in the same gauge theory. Notice that as far as fluxes are concerned,
this shift corresponds to moving one step down in the cascade. That this is the case, will become
apparent when we discuss the type IIA T-dual description of this system in the
next subsection.

Due to the $F_5$ background in the gravity solution (dual to the large number of D3
branes present in the cascade), the anti-D3 branes are attracted to the tip of the geometry.
The metastable configuration presumably has all anti-D3s absorbed and dissolved as gauge
flux into the $N_c$ D5 branes
\be
\int_{{\cal C}} {\cal F} = - N_c~,
\ee
where ${\cal C}$ is the 2-cycle which the $N_c$ D5s wrap. This flux, via
the Chern-Simons coupling in
the DBI action of the D5 branes, accounts for the $N_c$ units of anti-brane charge.
Stability against decay through the Myers
effect can be argued as in \cite{Argurio:2006ny}, but of course a more detailed
study would be valuable.

A natural question is to ask whether one of the anti-branes can annihilate with the deformation brane associated to node 4,
which is sitting at the conifold singularity. In the next subsection we will provide a simple
argument as to why this is energetically disfavoured.

The supersymmetric vacua corresponding to the baryonic branch were
discussed in \cite{Argurio:2006ny}. It is not immediately clear how one
would directly relate them to the metastable states.

\subsection{Type IIA dual}
\label{IIadual}

In this section we study the Type IIA dual configurations describing our model. These constructions were first introduced in \cite{Uranga:1998vf}.
This approach provides a vivid picture of how the anti-branes arise in
the non-supersymmetric state. 

\begin{figure}[ht]
  \centering
  \includegraphics[width=11cm]{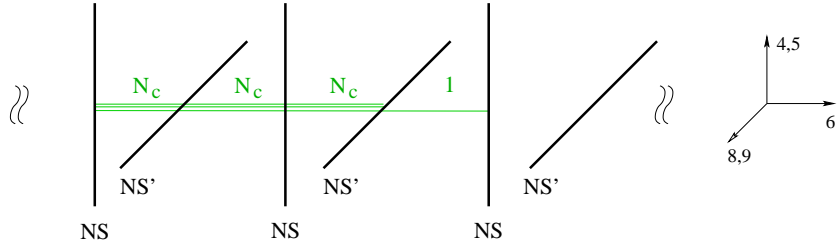}
  \caption{Type IIA configuration for the electric theory with
$\langle {M}_{22}\rangle=m=0$.
}
  \label{IIA_5nodes_1}
\end{figure}

\fref{IIA_5nodes_1} shows the type IIA T-dual brane configuration for our 4 node quiver for
$\langle {M}_{22}\rangle=m=0$.

\begin{figure}[ht]
  \centering
  \includegraphics[width=11cm]{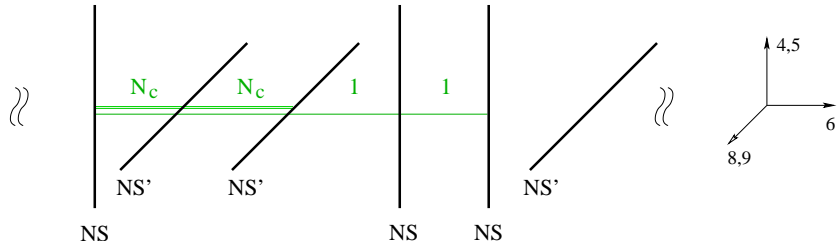}
  \caption{Type IIA configuration for the magnetic theory.}
  \label{IIA_5nodes_2}
\end{figure}

Performing a Seiberg duality on the middle node corresponds to moving the
second NS brane and second NS' brane across each other \cite{Elitzur:1997fh}. In the process,
some anti D4 branes are generated in the middle interval, but they are annihilated against
D4 branes sitting on top of them. The result is shown in \fref{IIA_5nodes_2}.

\begin{figure}[ht]
  \centering
  \psfrag{M22}{$\langle {M}_{22}\rangle$}
   \includegraphics[width=11cm]{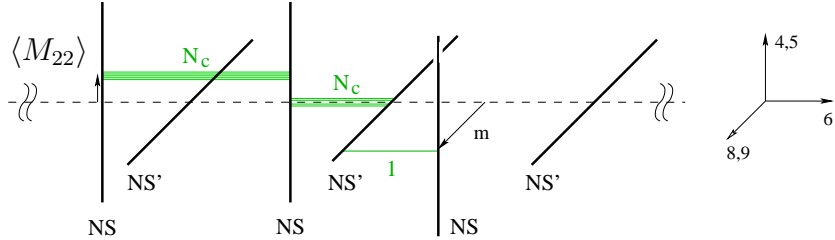}
  \caption{Type IIA configuration for the electric theory with
non-zero $\langle {M}_{22}\rangle$ and $m$.}
  \label{IIA_5nodes_3}
\end{figure}

A non-zero VEV for $\langle {M}_{22}\rangle$ in the electric theory corresponds to moving the D4-branes
stretched between the first and second NS-brane in \fref{IIA_5nodes_1} in the $45$ directions.
Similarly, the mass $m$ is mapped to a displacement of the third NS brane together with the D4-branes
that stretch from it to the second NS'-branes in $89$. This is shown in \fref{IIA_5nodes_3}.

\begin{figure}[ht]
  \centering
  \psfrag{M22}{$\langle {M}_{22}\rangle$}
  \includegraphics[width=11cm]{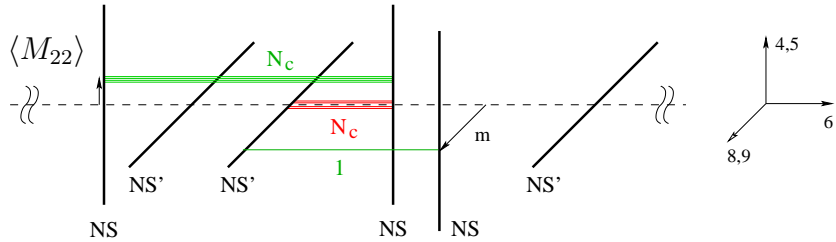}
  \caption{Type IIA configuration for the magnetic theory with
non-zero $\langle {M}_{22}\rangle$ and $m$. Anti-D4 branes are indicated in red.}
  \label{IIA_5nodes_4}
\end{figure}

If we now perform the Seiberg duality, we obtain the configuration in \fref{IIA_5nodes_4}.
The anti-D4's are not annihilated, since they are now displaced from the D4s due to the
meson VEVs (following the discussion in \S3, $\langle {M}_{22}\rangle$
is stabilized at tree level).

\fref{IIA_5nodes_4} shows that the avatar of SUSY breaking in the ISS vacuum
is explicit un-annihilated anti D4 branes.
The failure to annihilate these branes is a direct consequence of the meson VEVs and $m$ (the dynamically generated masses).
Relative to the IIB story we described in the previous subsection, this
IIA configuration is an intermediate picture between the state with
explicit anti-D3 branes and the (final) state where they are dissolved within the $N_c$ D5s
at the $A_1$ singularity.

\begin{figure}[ht]
  \centering
  \includegraphics[width=11cm]{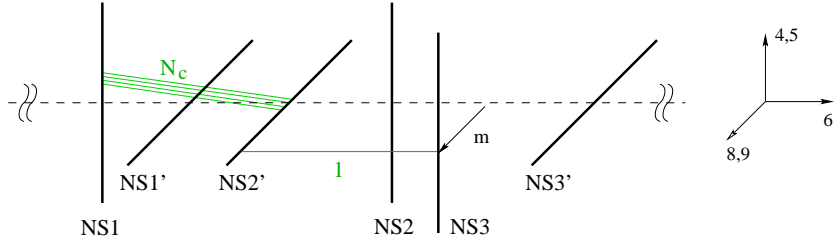}
  \caption{The type IIA configuration once the anti-D4 branes has been dissolved into the D4's as gauge flux.}
  \label{IIA_5nodes_8}
\end{figure}

Collapsing D4 and anti D4-branes as much as possible, we
obtain \fref{IIA_5nodes_8} for the IIA picture of the ISS vacuum. We have labeled the NS branes to
simplify the discussion.  We have annihilated the anti-D4's against D4's between NS1 and NS2 (and not NS2' and NS3) because,
since $|h\langle {M}_{22}\rangle| < m$, this clearly produces a lower energy configuration. Notice
that the objects the gauge flux (represented here by the anti-D4's) combines with are precisely D4 branes stretched between
the parallel NS1 and NS2. The resulting tilted D4 branes are dual to the D5/anti-D3 bound states
of the IIB configuration.

\begin{figure}[ht]
  \centering
  \includegraphics[width=11cm]{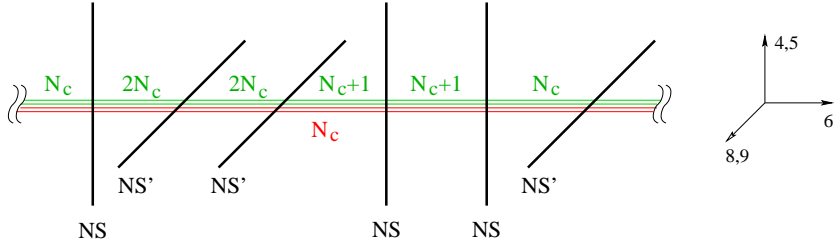}
  \caption{The ISS vacuum configuration of \fref{IIA_5nodes_4} can be interpreted as adding $N_c$ anti-D3 branes
  and moving down one step in the cascade.}
  \label{IIA_5nodes_6}
\end{figure}

In this type IIA setting it is also easy to see how
the configuration in \fref{IIA_5nodes_4} is related to the addition of anti D3-branes
and jumping fluxes in the gravity dual, as described in \S\ref{sugra}.
An anti-D3 brane maps to an anti-D4 brane
wrapping the entire $x_6$ circle in Type IIA. We can form such a complete anti-D4 by adding
D4/anti-D4 pairs to all intervals with the exception of the one between the second NS' and second NS. Grouping D4 branes in
each interval together, we are left with the configuration in \fref{IIA_5nodes_6}. It corresponds to $N_c$ full
anti-D4 branes (T-dual to $N_c$ anti-D3 branes) and the number of D4 branes in each interval corresponds
to moving up one step in the cascade from the magnetic theory (the last step). Increasing the cascade by one step is exactly how
increasing the NS flux by one unit manifests in this context. This matches nicely with our previous type IIB description
of the metastable non-supersymmetric vacua.

\section{Conclusions}

We have presented a D-brane construction that engineers metastable vacua closely related to those
of \cite{ISS}.  The construction has some interesting conceptual features and some interesting
features for model building.

Conceptually, the most interesting thing about the construction is that it readily admits
a IIB gravity dual description (in the general framework of AdS/CFT).  The metastable
states cannot be directly followed from weak 't Hooft coupling to strong 't Hooft
coupling, but they quite plausibly match on to strong-coupling analogues where the
SUSY breaking is well described by the presence of anti-D3 branes, and a
picture very similar to the one in \cite{KachruPearsonVerlinde}.  In addition, the quivers that arise
in our construction are some of the simplest cases where the new string instanton effects
explored in many recent works make important contributions.

One may also wish to construct pseudo-realistic models of SUSY breaking
and/or direct mediation using quiver
gauge theories.  In this case, our model has two virtues: the small dynamical masses
of the ISS model are explained naturally without any fine tuning of parameters
(as could also be done by retrofitting \cite{Retro}), and
the problematic R-symmetry which could
forbid gaugino masses is lifted by the extra
terms that automatically appear in our superpotential.
We note that because the R-symmetry is broken by an irrelevant operator
suppressed by the high scale $M_s^*$, and because for metastability it
is necessary to take $M_s^*$ somewhat higher than the SUSY-breaking scale,
it is likely that in real model-building applications, our theory would
produce low gaugino masses.  Given the lower bounds on gaugino masses,
this would necessitate heavy squarks and sleptons, resulting in the need
for a (mild) tune to obtain a reasonable Higgs mass. 

It 
would be very interesting to study the gravity dual geometry in further detail.
While there aren't BPS or protected quantities that are guaranteed to match between weak
and strong coupling, one may find interesting patterns of qualitative agreement between
the two classes of non-supersymmetric states. Conversely, some quantities (e.g. lifetimes
or barrier heights) may change in a striking way upon extrapolation in $g_sN$.

\begin{center}
\bf{Acknowledgements}
\end{center}
\medskip
We would like to
thank O. Aharony, F. Bigazzi, M. Buican, G. Ferretti, B. Florea, A. Lerda,
N. Saulina, N. Seiberg and A. Uranga for helpful discussions.
R.A. and M.B. are partially supported by the European Commission FP6
Programme MRTN-CT-2004-005104, in which R.A is associated to V.U. Brussel.
R.A. is a Research Associate of the Fonds National de la Recherche
Scientifique (Belgium). The research of R.A. is also supported by IISN - Belgium
(convention 4.4505.86) and by the ``Interuniversity Attraction Poles Programme --Belgian Science Policy''.
M.B. is also supported by Italian MIUR under contract PRIN-2005023102 and by a
MIUR fellowship within the program ``Rientro dei Cervelli''.
S.F. is supported by the DOE under contract DE-FG02-91ER-40671.
The research of S.K. was supported in part by a David and Lucile Packard Foundation
Fellowship, the NSF under grant PHY-0244728, and the DOE under contract
DE-AC03-76SF00515. S. F. would like to thank the Galileo
Galilei Institute for Theoretical Physics
for hospitality while this work was being completed.
S.K. similarly acknowledges the kind hospitality of the International Centre
for Theoretical Physics.

\appendix

\section{Fermionic zero modes and orientifolds}

\label{appendix_orientifold}

We now briefly explain how it is possible to project out two fermionic
zero modes on each instanton by embedding our model in an orientifold.

The most intuitive way of visualizing the orientifold is by means
of the Type IIA T-dual setup, along the lines of \cite{Park:1999ep}, to which we refer
the reader for further details. Everything in this construction can be mapped into a type IIB set-up.

\begin{figure}[ht]
  \centering
  \includegraphics[width=7.9cm]{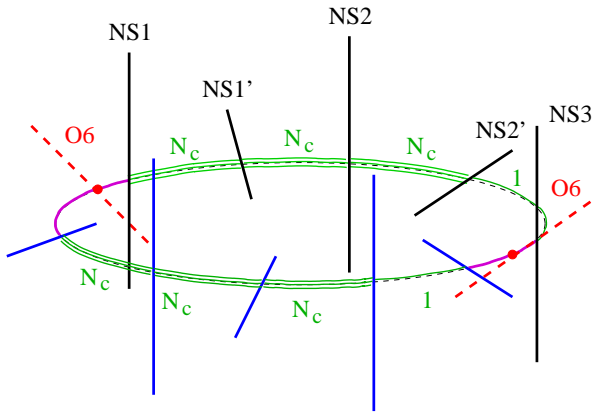}
  \caption{Type IIA T-dual configuration for an embedding of our model in an orientifold of 
a $\IZ_5$ orbifold of the conifold. The image NS-branes are indicated in blue. The ED1 branes are T-dualized to ED0s, which are shown in magenta.}
  \label{ori_inst}
\end{figure}

The simplest embedding of our model is shown in \fref{ori_inst}. It corresponds to
removing the last NS' in \fref{IIA_5nodes_1} and adding the orientifold images.
The O-plane extends along $01237$ and is at $45$ degrees with respect to the $45$ and $89$ planes.
Such an O-plane maps NS to NS' branes and vice versa. Before orientifolding, the
corresponding geometry is a $\IZ_5$ orbifold of the conifold.

\vskip -8pt

If the O-plane
extended along $45$ or $89$, the images of NS1 and NS3 would also be NS branes.
We have not chosen this possibility because the instantons stretched between
parallel NS-branes (in type IIB, this corresponds to
the instantons wrapping $\IP^1$'s in $A_1$ singularities) would have additional
adjoint fermionic zero modes, some of which also survive the orientifold projection.

We take the charge of the O-plane to be positive so that the gauge group on the instantons
is $O(1)$. Two of the fermionic zero modes are projected to the symmetric representation
and the other two to the antisymmetric representation
of $O(1)$. The antisymmetric representation of $O(1)$ vanishes, so this orientifold
projects out precisely two fermionic zero modes.

Only the cycles wrapped by the instantons are affected by the orientifold, thus our discussion of the IR applies without changes.
The cascade and the corresponding supergravity solution \S \ref{sugra}  require some small modifications in the presence of orthogonal gauge groups/orientifold planes, probably along lines similar to \cite{Imai:2001cq}.

\end{document}